\documentstyle[preprint]{jpsj}

\newcommand{\non}{\nonumber}

\newcommand{\e}{\varepsilon}
\newcommand{\ii}{{\rm i}}

\newcommand{\qpara}{q_{a}}
\newcommand{\qperp}{q_{b}}
\newcommand{\dd}{{\rm d}}

\newcommand{\beq}{\begin{eqnarray}}
\newcommand{\eeq}{\end{eqnarray}}
\newcommand{\k}{\mbox{\boldmath$k$}}

\newcommand{\be}{\begin{eqnarray}}
\newcommand{\en}{\end{eqnarray}}

\newcommand{\s}{\mbox{\boldmath$\sigma$}}
\newcommand{\ds}{\displaystyle}

\newcommand{\Q}{\mbox{\boldmath$Q$}}

\newcommand{\SDW}{\mbox{\boldmath$S$}}
\newcommand{\JPSJ}{J. Phys. Soc. Jpn.}

\newcommand{\PRL}{ Phys. Rev. Lett.}
\newcommand{\PRB}{ Phys. Rev. {\bf  B}}

\newcommand{\PROG}{Prog. Theor. Phys.}
\def\runtitle{Spin-Density-Wave Phase Transitions in  Quasi-One-Dimensional
Dimerized Quarter-Filled  Organic Conductors}
\def\runauthor{Jun-ichiro Kishine and Kenji Yonemitsu}
\title{Spin-Density-Wave Phase Transitions in  Quasi-One-Dimensional
Dimerized Quarter-Filled  Organic Conductors}
\author{Jun-ichiro Kishine\thanks{E-mail:kishine@ims.ac.jp} and Kenji Yonemitsu}
\inst
{Department of Theoretical Studies,
Institute for Molecular Science,
Okazaki 444-8585, Japan}
\recdate{February 25, 1999,\\
To appear in J. Phys. Soc. Jpn. {\bf 68} (1999) No.8}
\abst
{We have studied   spin density wave (SDW) phase transitions in   dimerized quarter-filled Hubbard chains
weakly coupled via   interchain one-particle  hopping, $t_{b0}$.
It is shown that there exists a   critical value of $t_{b0}$, $t_{b}^\ast$, between the
incoherent metal regime ($t_{b0}<t_{b}^\ast$) and the Fermi liquid regime ($t_{b0}>t_{b}^\ast$) in the metallic phase
above the SDW transition temperature.
By using  the 2-loop perturbative renormalization-group (PRG) approach 
 together with the random-phase-approximation,
  we propose a SDW phase diagram  covering both of the  regimes.
The SDW phase transition from the incoherent metal phase for $t_{b0}<t_{b}^\ast$
is caused by
growth of the intrachain electron-electron umklapp scattering toward low temperatures, which
is regarded as  
 preformation of the Mott gap.
We discuss relevance of the present result to the SDW phase transitions in the
quasi-one-dimensional dimerized quarter-filled organic conductors, (TMTTF)$_2$X and (TMTSF)$_2$X.}
\kword
{ dimerization, umklapp scattering,  interchain hopping,  incoherent metal,  
Fermi liquid, spin density wave,  perturbative renormalization-group}

\begin{document}
\maketitle

\baselineskip 16pt

\section{Introduction}

Interplay of quantum fluctuations and   dimensionality effects in  quasi-one-dimensional (Q1D)
organic conductors (TMTTF)$_2$X and (TMTSF)$_2$X (X=Br, PF$_6$,...) has been extensively
studied from both experimental and theoretical sides over the past two decades.\cite{IshiguroSaitoYamaji98}
Especially,  different nature of  the  metallic phases above the spin-density-wave transition temperatures of
(TMTTF)$_2$Br and (TMTSF)$_2$PF$_6$ at ambient pressure
has provoked a great deal of controversy.
These compounds, in common, consist of   quarter-filled  
chains with  dimerized
one-particle hopping integrals along the chain ($a$-axis), $t_{a1}$ and $t_{a2}$, 
and an interchain one-particle hopping integral,  $t_b$, along the intermediately conducting axis ($b$-axis).
Hopping integrals  along the third direction ($c$-axis) in both compounds are about one tenth of $t_b$.

Direct evidence of different nature of the metallic phases in   the TMTTF and TMTSF compounds
has been given by  
the optical reflectivity measurements  which indicate  the one-particle motion is nearly confined into the chains
in (TMTTF)$_2$Br, while deconfined in (TMTSF)$_2$PF$_6$.\cite{Jacobsen81,Vescoli98}
Temperature dependence of the $a$-axis resistivity in   (TMTTF)$_2$Br shows 
a shallow minimum around the so-called charge localization temperature and
 has been interpreted as
preformation of a charge localization gap (Mott gap).\cite{Emery82}
On the other hand, it has been widely accepted that, as temperature decreases, 
  (TMTSF)$_2$PF$_6$ undergoes a crossover to a Fermi liquid (FL) regime
and the SDW phase transition in (TMTSF)$_2$PF$_6$ is driven by the Fermi surface nesting,\cite{Yamaji82,HF86,Shimahara88} below the crossover temperature.
Although experimental assignment of the crossover temperature  is still highly controversial,\cite{Moser98,Schwartz98,Fertey99}
existence of the robust Fermi surface  in the TMTSF compounds has  also been    supported through   
the success of the explanation of 
the field-induced SDW phenomena in the Fermiology scheme.\cite{GorkovLebed,Yamaji85,Osada}
Dimensionality effects on 
the SDW phase transition  was also studied
in (TMTTF)$_2$Br under pressure $P$, indicating    that $T_N$ 
  increases for   $P<P_{\rm opt}=5$kbar, while decreases    
 for  $P>P_{\rm opt}$
and the crossover between the two regimes is roughly coincident with vanishing of the
charge localization temperature.\cite{Klemme96,Takahashi98}
Similar pressure dependence was also observed in (TMTTF)$_2$PF$_6$.\cite{Moser98}

Q1D conductors behave as  1D systems at high temperatures,
 $T \gg t_{b0}$, where $t_{b0}$ is a bare interchain one-particle hopping integral.
Low-energy asymptotic behavior of 1D systems 
 has been well understood by way of the renormalization-group (RG) approach based on the
scaling hypothesis.\cite{Solyom}
In the absence of $t_{b0}$, due to the electron-electron umklapp scattering, the dimerized 
quarter-filled chain is scaled to a Mott insulator,\cite{PencMila94} 
which is characterized by the  Mott gap  and the antiferromagnetic power-law 
correlation.
In the presence of small $t_{b0}$,   as the temperature decreases,   
there occur  the interchain one-particle propagation through the $t_{b0}$ process
and the propagation of the 1D antiferromagnetic (1DAF)   power-law correlation 
through
the interchain   particle-hole  exchange (ICEX) processes.  
The former process  drives the crossover to the FL regime, while the latter process
 converts the 1DAF power-law correlation to the 2D (or 3D) long-range correlation.
Since the latter process occurs {\it irrespective of the interchain quasiparticle coherence},
a phase transition from an \lq\lq incoherent metal phase\rq\rq  \, occurs, if the latter dominates the former.
The ICEX mechanism of   phase transitions in   Q1D systems was pointed out by Brasovskii and Yakovenko,\cite{BY86}
and confirmed by Suzumura.\cite{Suzumura85}

So far, importance of the umklapp scattering and the resultant Mott gap in the TMTTF compounds has been extensively
 discussed,\cite{BourbonnaisCaron91,Bourbonnais93a,Bourbonnais93b,Bourbonnais97,
Giamarchi97,Suzumura98,Tsuchiizu99a,Tsuchiizu99b} in terms of   the weak-coupling RG approach.
Bourbonnais\cite{Bourbonnais97} argued that  in the case of the TMTTF 
a coherent interchain one-particle hopping is prohibited due to  formation of the Mott gap.
Consequently the interchain AF interaction strength is given by $J_\perp\sim v_F \tilde t_b^2/\Delta_{\rho}^2$,
where $\tilde t_b$  is  a renormalized interchain one-particle hopping integral 
and $\Delta_{\rho}$ is the preformed Mott gap 
which was introduced phenomenologically.
Then the Stoner criteria gives the N$\acute{\rm e}$el temperature, $T_{\rm N}\sim {\tilde t_b}^2/\Delta_{\rho}$, 
which increases when $\tilde t_b$ increases or  $\Delta_{\rho}$ decreases  
under pressure. 
Based on the bosonization approach, Suzumura {\it et al}. \cite{Suzumura98,Tsuchiizu99a,Tsuchiizu99b}
discussed  a confinement-deconfinement transition
  in the half-filled two coupled  chains 
in terms of  a misfit parameter due to   the interchain one-particle hopping integral and  the  Mott gap
of the  isolated half-filled chain at $T=0$.

In the present paper, 
we extend the work  shortly presented previously\cite{JK98ICSM}  and 
 discuss the SDW phase transitions from  both the incoherent metal and the FL regimes 
 in the  dimerized quarter-filled Hubbard chains weakly-coupled via the   interchain one-particle hopping,
$t_{b0}$.
Based on the 2-loop RG approach,
we show that there exists a   critical value of $t_{b0}$, $t_{b}^\ast$, between the
incoherent metal regime ($t_{b0}<t_{b}^\ast$) and the Fermi liquid regime ($t_{b0}>t_{b}^\ast$) in the metallic phase
above the SDW transition temperature.
We assume that the scaling hypothesis in the 1D regime works well down to the SDW phase transition temperature 
for $t_{b0}<t_{b}^\ast$
and determine the  transition temperature, $T_{\rm N}^{\rm RG}$,   based solely on  
the 2-loop renormalization-group flows, without introducing the preformed Mott gap
 phenomenologically.
In  the present framework,
 {\it growth of the umklapp scattering toward low temperatures plays an essential role on  occurrence of
the SDW phase transition from the incoherent metal phase}.
As we shall discuss in \S 5, growth of the umklapp scattering strength is regarded as  
 the preformation of the Mott gap
and thus our results are consistent with the views based on the Mott 
gap.\cite{BourbonnaisCaron91,Bourbonnais93a,Bourbonnais93b,Bourbonnais97,
Giamarchi97,Suzumura98,Tsuchiizu99a,Tsuchiizu99b}

Outline of the present paper is as follows. 
In \S 2,  we give a full account of  the 2-loop RG treatment  for the intra-  and inter-chain processes and
 determine  $T_{\rm N}^{\rm RG}$   and $t_{b}^\ast$.  
In \S 3,  the nesting-driven SDW phase transition in the FL regime   is discussed based on the random-phase-approximation (RPA).
In \S 4, we give a  phase diagram covering   both of the   regimes,  $t_{b0}<t_{b}^\ast$ and $t_{b0}>t_{b}^\ast$.
In \S 5, we  
  discuss relevance of the present work
to the views based on the Mott 
gap. We also comment on the problems which remain unsettled in the present scheme.
Finally we conclude our work.
 In this work,  we concentrate on  the   SDW phase transitions 
 and do not discuss  
the spin-fluctuation-mediated superconducting phase transition which has  been   extensively
discussed by several authors.\cite{HF86,Shimahara88,Yamaji83}

\section{SDW Phase Transition from the  Incoherent Metal Regime}
\subsection{Model and effective Hamiltonian}
We consider an array of a large number of   dimerized quarter-filled Hubbard chains   on the $a$-$b$s plane (Fig.~1).
The nearest neighbor chains are
weakly coupled via
the 
interchain one-particle hopping $t_{b0}$. The  dimerization along the  $a$-axis
and the  intrachain on-site Coulomb repulsion, $U$, are taken into account.
As shown in Fig.~2, due to the dimerization the  one-particle dispersion  is  split into the upper and lower bands  
 given by
\be
\pm\sqrt{t_{a1}^2+t_{a2}^2+2t_{a1}t_{a2}\cos k_a  }-2t_{b0}\cos k_b,\label{eqn:1Pdis}
\en
where electron wave numbers along the a- and b-axes are denoted by
$k_a$ and $k_b$, respectively. 
In  the temperature scale considered here, which is much smaller than the Fermi energy or intrachain one-particle hopping integrals,
 electron dynamics is confined into the effectively half-filled lower band.
In the absence of $t_{b0}$ and $U$, the  dispersion relation, (\ref{eqn:1Pdis}),
gives the density of states  per spin 
$
{\cal N}(\e)=
-{1\over 2\pi t_{a1}t_{a2} } 
\e/\sqrt{1- \{\e^2-t_{a1}^2-t_{a2}^2)/ 2t_{a1}t_{a2})\}^2},
$
which relates the
chemical potential, $\mu$,   to the carrier density, $n$ ($n=1$ corresponds to the  half-filled lower band), as
$
n=1-{2\over  \pi }\sin^{-1}[(t_{a1}^2+t_{a2}^2-\mu^2)/  2t_{a1}t_{a2} ] . \label{eqn:mod2Ddisper}
$
Thus,  
$n=1$ corresponds to the Fermi wave number, $k_F=\pi/2$, and the  Fermi velocity, 
$ 
v_F=t_{a1}t_{a2}/\sqrt{t_{a1}^2+t_{a2}^2}.
$
Taking account of    modification of the Fermi wave number due to the weak dispersion in the $b$-axis direction, as was done  in Refs.[6] and [11],
we obtain  the one-particle dispersion linearized in the $k_a$-direction at the Fermi points $k_F$ as
\be
\xi(\k)=v_F(\mid\!k_a\!\mid-k_F)- 2t_{b0} \cos k_b+2t_{b0}'\cos 2 k_b +{\cal O}(t_{b0}^3/t_{a1}^2) , \label{eqn:1Pdispersion}
\en
where
\be
t_{b0}'={ t_{b0}^2\over W} \sqrt{2\over 1+\delta^2},
\en
with
$W=2(t_{a1}+t_{a2})$ and $\delta=(t_{a1}-t_{a2})/(t_{a1}+t_{a2})$
being the total bandwidth and the dimerization ratio in the absence of $t_{b0}$ and $U$.
The third term of (\ref{eqn:1Pdispersion}),   breaks  the perfect nesting of the Fermi surface.

In the absence of $t_{b0}$,   
the one-particle thermal coherence length   along the $a$-axis is given by 
$\xi=v_F/T$ with 
 $T$ being  the temperature,\cite{comment0} 
and it  becomes  
 much larger than the intrachain lattice spacing at temperatures, $T\ll v_F$.
 This situation enables us  to
  take the continuum limit along the $a$-direction and to apply the renormalization-group analysis to the intrachain system based on the
{\it scaling hypothesis} in the 1D regime. 
On the other hand, since  the temperature scale considered here can become comparable with the small  perturbation, $t_{b0}$,
 the one-particle  thermal coherence length  along the $b$-axis becomes comparable with  
the  distance between the adjacent chains. Thus {\it  we must keep lattice discreteness along the $b$-axis.}

Based on the bandwidth regularization scheme, as shown in Fig.~2, we restrict the electron wave numbers {\it along the $a$-axis}   to the regions 
\be
{\cal C}_l=\{k_a \mid -\omega_l/2\leq \xi_{\nu}(k_a) \leq \omega_l/2\},
\en
 where $\xi_{R}(k_a)=v_{F}(k_a-k_{F})$ ($k_a>0$) and
$\xi_{L}(k_a)=v_{F}(-k_a-k_{F})$ ($k_a<0$) are the linearized  dispersions for the right-  and left-moving electrons.
The cutoff of the linearized band is 
parameterized  as 
$
\omega_l=E_{0}{\rm e}^{-l}
$
 with the scaling parameter, $l$. 
 The cutoff energy   $\omega_l$ corresponds to a characteristic  energy   at which we observe the system.
From now on  we regard $\omega_l$ as the temperature scale   
$
\omega_l\sim T.
$
As  $l$ goes 
from zero to infinity, we move from
  high-temperature scales, where the system is regarded as the    1D chains, to   low-temperature scales where the interchain couplings 
play  important roles.

We start with the    effective Hamiltonian   which depends on the energy scale, $l$, as
\be
{\cal H}_{l}={\cal H}_{a;l}^{(1)}+{\cal H}_{a;l}^{(2)}+{\cal H}_{b;l}^{(1)}+{\cal H}_{b;l}^{(2)},\label{eqn:Hamiltonian}
\en
which  are described below.
The    intrachain one-particle term is written as
\begin{eqnarray}
{\cal H}^{(1)}_{a;l}=\sum_{i=1}^{N_b} \sum_{k_{a}\in {\cal C}_l}  
\sum_{\sigma}
\left[\xi_{R} (k_a)
R^{\ast}_{i,\sigma}(k_a)R_{i,\sigma}(k_a)
+\xi_{L} (k_a)
L^{\ast}_{i,\sigma}(k_a)L_{i,\sigma}(k_a)\right],\label{eqn:intra1P}
\end{eqnarray}
where
$R_{i,\sigma}$ ($R_{i,\sigma}^\ast$) and $L_{i,\sigma}$ ($L_{i,\sigma}^\ast$) are  
fermion annihilation (creation) operators  representing the right- and left-moving electrons with spin $\sigma$, respectively, on the $i$-th chain in the vicinity of the Fermi 
points in the lower band, and $N_b$ is the number of the chains. 
As will be discussed in \S 2.2 and the Appendix, the scale-invariance under the renormalization transformation in the 1D regime is  imposed 
on ${\cal H}^{(1)}_{a;l}$.

The interchain one-particle process [Fig.~3 (a)] is renormalized through the intrachain self-energy effects,\cite{BourbonnaisCaron91}
and consequently  become dependent  on the energy-scale.
By introducing the Fourier transform, $R_{i,\sigma}(k_a)={1\over \sqrt{N_b}}\sum_{k_b}e^{\ii k_b i}R_\sigma(\k)$  with $\k=(k_a,k_b)$,
and $L_{i,\sigma}(k_a)$ likewise,
the interchain one-particle term is written as
\begin{eqnarray}
{\cal H}^{(1)}_{b;l}=-2\sum_{k_a\in {\cal C}_l}\sum_{-\pi\leq k_b\leq \pi} \sum_{\sigma}  (t_{b;l}\cos k_b
-t_{b;l}'\cos 2k_b)
\left[
 R^{\ast}_{\sigma}(\k)R_{\sigma}(\k)
+ L^{\ast}_{\sigma}(\k)L_{\sigma}(\k) 
\right].\label{eqn:inter1P}
\end{eqnarray}
Unrenormalized values of $t_{b;l}$ and $t_{b;l}'$ are denoted  by
$
t_{b0}
$
and
$
t_{b0}'
$,
respectively.

The   intrachain two-particle scattering processes contain the normal  [Fig.~3(b),(c)]
and     umklapp  [Fig.~3(d)] processes with
the  dimensionless scattering strengths, $g_{l}^{\sigma_1\sigma_2\sigma_3\sigma_4}$ and $g_{3;l}$, respectively.\cite{Solyom} 
The corresponding term is written as
\begin{eqnarray}
{\cal H}^{(2)}_{a;l} &=&
{\pi v_{F} \over N_a } \sum_{i=1}^{N_b}\sum_{k_{ai}\in C_l}\sum_{\sigma_i}g_{l}^{\sigma_1\sigma_2\sigma_3\sigma_4}
R^{\ast}_{i,\sigma_1}(k_{a1})L^{\ast}_{i,\sigma_2}(k_{a 2}) 
L_{i,\sigma_3}(k_{a3}) R_{i,\sigma_4}(k_{a 4})  \non\\ 
&+&
{\pi v_{F} \over 2 N_a }g_{3;l} \sum_{i}^{N_b}\sum_{k_{ai}\in C_l}\sum_{\sigma,\sigma'}
\left[
R^{\ast}_{i,\sigma}(k_{a 1})R^{\ast}_{i, \sigma'}(k_{a2}) 
L_{i, \sigma' }(k_{a3}) L_{i,\sigma }(k_{a 4})+{\rm c.c}\right] , \label{eqn:2pint}
\end{eqnarray}
where $N_a$ denotes the number of sites in the chain.
The summation over momentum  is taken under the constraint:
$k_{a1}+k_{a2}-k_{a3}-k_{a4}=G$ with $G=0$ and $G=\pm 4k_F=\pm 2\pi$ for  the normal and umklapp processes, respectively.
The normal scattering  is decomposed into backward [Fig.~3(b)] and  forward [Fig.~3(c)]  scattering  as
\be
g_{l}^{\sigma_1\sigma_2\sigma_3\sigma_4}=\delta_{\sigma_1\sigma_4}\delta_{\sigma_2\sigma_3}g_{2;l}
-\delta_{\sigma_1\sigma_3}\delta_{\sigma_2\sigma_4}g_{1;l},
\en
where the forward and backward scattering strengths are denoted by $g_{2;l}$
and $g_{1;l}$, respectively.
Unrenormalized scattering strengths   are related to the on-site Coulomb repulsion,
 $U$, as\cite{PencMila94}
\begin{eqnarray}
\pi v_F g_{1;0}=\ds{U\over 2} ,\,\,\,\,\,\,
\pi v_F g_{2;0}=\ds{U\over 2},\,\,\,\,\,\,
\pi v_F g_{3;0}=\ds{U\over 2}\ds{2 \delta\over 1+\delta^2}.\label{eqn:Uvsg}
\end{eqnarray}
The umklapp scattering strength depends on the dimerization ratio,\cite{PencMila94} in contrast with  the half-filled case.

By multiple use of the interchain one-particle hopping and the intrachain two-particle interaction,
the ICEX processes are dynamically generated during the renormalization.\cite{BY86,BourbonnaisCaron91}
As described below,   we consider only the case where the most dominant 1D power-law correlation   is an   antiferromagnetic  one.  
In this case, as in the  weakly-coupled half-filled chain system,\cite{JK98}
the most dominant ICEX   process is in the AF channel.
The corresponding term 
is written as
\be
{\cal H}_{b;l}^{(2)}&=&{\pi v_{F}\over   N_a }\sum_{i\neq j}\sum_{q_a} 
J_{ i-j ;l}
{\SDW}_{i;l}(\qpara )\cdot{\SDW}_{j;l}^\ast(\qpara )\label{eqn:Hb2} \\
&+&{\pi v_{F}\over   N_a } \sum_{i\neq j}\sum_{q_a} 
K_{ i-j;l }\left[
{\SDW}_{i;l} (2k_F+\qpara )\cdot{\SDW}_{j;l} (2k_F-\qpara )
+{\rm c.c}\right]\non,
\en
where
 $J_{i-j;l}$  and $K_{i-j:l}$   represent 
 the strengths of  interaction  
      between the 
$i$-th and $j$-th chains through
the normal and umklapp scattering, respectively [see Fig.~3(e) and 3(f)]. 
It should  be noted that, in the presence of the intrachain umklapp process, we have to include
the $K_{ i-j;l }$-process which was not taken into account in Ref.[21].
The $2k_F$ spin-density on the $i$-th chain is given by
\be
{\SDW}_{i;l}(\qpara )= \sum_{\stackrel{\scriptstyle k_a\in C_l}{k_a+q_a\in C_l}} R^\ast_{i,\alpha}(k_a+q_a )
{\s_{\alpha\beta}\over 2 }L_{i,\beta}(k_a   ).
\en
The interaction strengths are initially zero,  
\be
J_{i-j;0}=K_{i-j;0}=0,
\en
but   dynamically generated during the renormalization process.
By taking the Fourier transforms,
$
 J_{q_b;l}={1\over  N_b }\sum_{i\neq j}e^{\ii q_b   (i-j)}J_{i-j;l}
$, 
$
 K_{q_b;l}={1\over  N_b }\sum_{i\neq j}e^{\ii q_b   (i-j)}K_{i-j:l}
$
and
$
{\SDW}_l(q_a,q_b )={1\over  N_b }\sum_{q_b}e^{\ii q_b    i }{\SDW}_{i;l}(q_a),
$
the term, (\ref{eqn:Hb2}), is rewritten as
\be
{\cal H}_{b;l}^{(2)} &=&{\pi v_{F}\over  N_a N_b} \sum_{q_a,q_b} 
J_{ q_b ;l}
{\SDW}_l^\ast (q_a,q_b )\cdot{\SDW}_l(q_a,q_b ) \\
&+&{\pi v_{F}\over  N_a N_b} \sum_{q_a,q_b} 
K_{q_b;l }\left[
{\SDW}_l  (2k_F+\qpara,q_b)\cdot{\SDW}_l (2k_F-\qpara,-q_b)
+{\rm c.c}\right]\non.
\en

\subsection{Renormalization-group equations}
The PRG approach\cite{BourbonnaisCaron91} is based on the assumption  that    $g_{i;l}$ and $t_{b;l}$  
are
considerably smaller than $E_0$  and sets 
up   low-order scaling equations whose solutions indicate whether these small perturbations grow toward the low-energy scales  or not. 
During this step, we  take account of the  intrachain scattering and self-energy processes at the 2-loop level.
Outline of the derivation of  the   RG equations  
is left to   the Appendix.

Renormalization of the interchain one-particle hopping comes solely from the intrachain self-energy processes and
the corresponding RG equations are given by\cite{BourbonnaisCaron91,Bourbonnais93a,Bourbonnais93b}
\be
{\dd \over \dd l} \ln t_{b;l}&=&1-\theta_l,\label{eqn:rgeqt1}\\
{\dd \over \dd l} \ln t_{b;l}'&=&1-\theta_l,\label{eqn:rgeqt2}
\en
where a non-universal exponent $\theta_l$ comes   from  the intrachain self-energy processes as  shown   in  Fig.~4
and is given by\cite{Kimura75,Bourbonnais93a} 
\be
\theta_l={1\over 4} \left[g_{1;l}^2+g_{2;l}^2-g_{1;l}g_{2;l}+ g_{3;l}^2/2\right].
\en
During the renormalization process, no new   interchain one-particle hopping is generated.
We see from eqs.~(\ref{eqn:rgeqt1}) and (\ref{eqn:rgeqt2}) that the ratio $t_{b;l}'/t_{b;l}$
is scale-invariant:
\be
t_{b;l}'/t_{b;l}=t_{b0}'/t_{b0}={ t_{b0}\over W} \sqrt{2\over 1+\delta^2}.\label{eqn:ratio}
\en 

The 2-loop RG equations for the intrachain  normal and umklapp scattering strengths are 
given by\cite{Kimura75,Solyom}
%%%%%%%%%%%%%%%%%%%%%%%%%%%%%%%%%%%%%%%%%%%%%%%%%%%%%%%%%%%%%%%%%%%%%%%%
\begin{eqnarray}
{\dd\over \dd l} g_{1;l}  &=&-{g_{1;l}} ^2-{1\over 2}{g_{1;l}} ^3 ,\label{eqn:rgeqg1}\\
{\dd \over \dd l}G_l   &=&-{g_{3;l}} ^2\left(1+{1\over 2}G_l \right),\\
{\dd \over \dd l} g_{3;l}   &=&-g_{3;l} G_l \left(1+{1\over 4}G_l  \right)-{1\over 4}{g_{3;l} }^{3} \label{eqn:rgeqg3},
\en
where $G_l=g_{1;l}-2g_{2;l}$.
The intrachain charge degrees of freedom  are governed by the combination $(G_l,g_{3;l})$ with  flow lines $(G_l-{\rm const.})^2-{g_{3;l}}^2={\rm const}$.
When the unrenormalized values, $G_0$ and $g_{3;0}$ satisfy the condition, 
\be
G_0< \mid\! g_{3;0}\! \mid , \label{eqn:condition}
\en
 the umklapp process becomes relevant
and the 2-loop RG eqs. (\ref{eqn:rgeqg1}) $\sim$ (\ref{eqn:rgeqg3}) give the non-trivial fixed point, $g_{1;\infty} =0$ 
and $\mid\! g_{3;\infty} \!\mid=-G_{\infty}=2$.\cite{Solyom}
In the absence of the interchain coupling,  the low-energy asymptotics  is  a Mott insulator   with the dominant AF 
power-law correlation.\cite{Solyom}
As far as the condition (\ref{eqn:condition}) is satisfied, all the results described below are  qualitatively unchanged.
Thus, from now on except in Figs.~8 and 13 shown below, we fix $U$  at
$U=1.45\pi v_F$ which corresponds to
$
g_{1;0}=g_{2;0}=0.725$  and  
$ 
g_{3;0}=0.725\times\ds{2 \delta/( 1+\delta^2)}.
$

We note that the non-trivial fixed point, $g_{1;\infty}=0$ and $\mid g_{3;\infty}\mid=-G_{\infty}=2$,  gives
\beq
{\dd\ln t_{b;l}/ \dd l}\stackrel{l\to\infty}{\longrightarrow}{1/ 4}.\label{eqn:FP}
\eeq
Thus, for large $l$, $t_{b;l}$ grows as $t_{b;l}=t_{b0}e^{l/4}$. 
Consequently,   $t_{b}$ is a {\it relevant} perturbation in the RG sence
and always attains an order of  the initial bandwidth, $E_{0}$, 
at some crossover value of the scaling parameter  qualitatively defined by
$
t_{b;l_{\rm cross}}= E_{0}. 
$
Here we   stress that  {\it the relevance or irrelevance of $t_{b;l}$ in the 
asymptotic limit of the RG flow
 makes no qualitative difference
on the crossover to the FL regime at a finite energy scale}.
Actually, the 1-loop RG analysis gives the strong coupling fixed point $\mid\! g_{3;\infty} \!\mid=-G_{\infty}=\infty$\cite{Solyom} and accordingly
    $t_{b;l}$ becomes irrelevant~[see eq.~(\ref{eqn:rgeqt1})].
In this case the flow of $\theta_l$ deviates from that obtained by the 2-loop analysis given here only at 
 energy scales much lower
than $T_{\rm cross}$ defined below by eq.~(\ref{eqn:Tcross}).
This is the very reason why it is important to {\it keep track of the whole RG flow of } $t_{b;l}$.

The temperature scale,
\be
T_{\rm cross}=E_0 {\rm e}^{-l_{\rm cross}}, \label{eqn:Tcross}
\en
gives
 a qualitative measure around which the interchain   one-particle motion begins to develop.\cite{Boies95}
 In Fig.~5, we show the RG flows of $t_{b;l}$ for various  dimerization ratios, $\delta=0,0.2,0.4,0.8$.
We see that    finite dimerization strongly suppresses the growth of $t_{b;l}$, since
the increasing $\delta$ causes the stronger umklapp scattering which suppresses more and more severely the 
interchain one-particle propagation.

In Figs.~6(a) and 6(b), we show contribution to the RG equations for the ICEX processes 
in the AF channel for
 the normal and umklapp scattering processes,
respectively. The respective equations  are given by
 \be
{\dd \over \dd l} J_{q_b;l}&=&{1\over 2}\left[{g_{2;l}} ^2+4{g_{3;l}}^2\right]
f_{l}(q_b)\label{eqn:rgeqJ}
 \\
&+&{1\over 2}\left[(g_{2;l}-4\theta_l) J_{ \qperp;l}+4g_{3;l}  K_{q_b;l } \right]
-{1\over 4}\left[J_{q_b;l}^2+4K_{q_b;l}^2\right],  \non\\
{\dd \over \dd l}K_{q_b;l} &=&2 
{g_{2;l}}{g_{3;l}}f_{l}(q_b)
+2\left[(g_{2;l}-\theta_l) K_{q_b;l} +g_{3;l} J_{q_b;l} \right]-J_{q_b;l}K_{q_b;l} ,\label{eqn:rgeqK}
\end{eqnarray}
%%%%%%%%%%%%%%%%%%%%%%%%%%%%%%%%%%%%%%%%%%%%%%%%%%%%%%%%%%%%%%%%%%%%%%%%
where   
\be
f_{l}(q_b)&=&{ \tilde t_{b;l} }^2\cos\qperp
+\tilde t_{b;l}'^2  \cos 2\qperp\non\\
&=&{ \tilde t_{b;l} }^2\left[\cos\qperp+\left({t_{b0}'\over t_{b0}}\right)^2\cos2\qperp\right]
\en
with $\tilde t_{b;l} = t_{b;l}/ E_0$ and $\tilde t_{b;l}' = t_{b;l}'/ E_0$.
Figs.~6(a-1) and 6(b-1) represent the dynamical generation of these processes and give the first terms of (\ref{eqn:rgeqJ}) and (\ref{eqn:rgeqK}).
Figs.~6(a-2) and 6(b-2) represent the coupling of the intrachain  AF process to
the interchain   process     and give the second terms of (\ref{eqn:rgeqJ}) and (\ref{eqn:rgeqK}).
Figs.~6(a-3) and 6(b-3) represent the coupling of the interchain    processes
  themselves and  give the third terms of (\ref{eqn:rgeqJ}) and (\ref{eqn:rgeqK}).

Although   the unrenormalized values of  $J_{q_b;0}$ and $K_{q_b;0}$ are zero,
the first terms of  eqs.(\ref{eqn:rgeqJ}) and (\ref{eqn:rgeqK}) generate finite magnitudes of 
$J_{q_b;l}$ and $K_{q_b;l}$, then   the second terms induce their exponential growth,
and  finally the third terms cause their divergence at the 
critical scaling parameter $l_{c}(q_b)$ which depends on the   momentum $q_b$  in the $b$-axis direction,
$
J_{q_b;l_c(q_b)}=K_{q_b;l_c(q_b)}=-\infty.
$
The divergence corresponds to the phase transition to the long-range ordered phase at 
temperature, $T_{\rm N}^{\rm RG}(q_b)=E_{0}e^{-l_c(q_b)}$.
At the most favorable SDW vector, 
$
f_{l}(q_b)
$
becomes negative and has a maximum absolute value.
This vector is always given by  
 $q_b=\pi$, when the small $t_{b0}$ satisfies
$
t_{b0}'/t_{b0}={ t_{b0}\over W} \sqrt{2\over 1+\delta^2}<1/2.
$
Thus, within the present scheme, the ICEX mechanism causes
the commensurate SDW state characterized by the SDW vector,
\be
\Q_{\rm RG}=(2k_F,\pi),
\en
with $k_F=\pi/2$.
From now on,  we fix $q_b=\pi$ in
$J_{q_b;l}$  and $K_{q_b;l}$  and introduce the SDW transition   temperature,  
\begin{eqnarray}
T_{\rm N}^{\rm RG}= E_{0}{\rm e}^{-l_{N}}, 
\end{eqnarray}
with $l_N=l_{c}(\pi)$.
In our formulation, contribution from the interchain processes  to the intrachain processes are not taken into account.
As a result, the present treatment of the phase transition is analogous to the interchain mean-field theory of  quasi-one-dimensional
systems.\cite{SIP}

As is seen from the diagrams in Figs.~6(a) and (b), the intrachain umklapp processes couple strongly to the ICEX processes.
Consequently, growth of the umklapp scattering toward low temperature strongly
  enhances the ICEX processes. On the other hand, as we previously noted, the umklapp scattering  suppresses the interchain
one-particle process. As a consequence of these combined effects, 
there appears a region where the SDW phase transition from the incoherent metal phase driven by the ICEX process dominates
the interchain  one-particle crossover to the FL regime. 
It is noted that, in the absence of the umklapp process, the one-particle crossover always 
dominates the ICEX-driven phase transition, provided $U$ is not extremely large.\cite{Boies95}

\subsection{AF phase transition temperature and the critical value of $t_{b0}$}
By solving the coupled RG equations (\ref{eqn:rgeqt1}), (\ref{eqn:rgeqg1}) $\sim$ (\ref{eqn:rgeqg3}), (\ref{eqn:rgeqJ}) and (\ref{eqn:rgeqK})
and   keeping track of the RG flows of
the interchain one-  and two-particle processes at finite energy scales,
 we    compare the growth of the interchain one-particle propagation  
through the $t_{b0}$ process with that of the interchain propagation of the dominant 1DAF correlation through the ICEX process.
When the RG equations give $T_N>T_{\rm cross}$,
  the SDW
phase transition 
driven by the ICEX mechanism occurs.
Otherwise,  evolution of 2D  quasiparticle coherence leads the system to the 2D  Fermi liquid (FL) regime.

In Figs.~7(a), 7(b), and 7(c) we show the RG flows of $g_{3;l}$, $t_{b;l}/E_{0}$, 
$J_{q_b;l}$ and $K_{q_b;l}$ with $q_b=\pi$ for  $\delta=0.2$ and
$ t_{b0}/E_0=0.02, 0.108,   0.3$,
where  the vertical  lines show the locations of $l_{\rm cross}$ and $l_{N}$.
RG flows of the umklapp scattering strength, $g_{3;l}$,  depend only on the intrachain Hubbard repulsion, $U$, and grow
 toward the
strong coupling fixed point, $g_{3;\infty}=2$.
We here assume that the scaling procedure works well down to $T_{\rm N}^{\rm RG}$, although $g_{3;l}$ exceeds unity at 
 energy scales
higher than   $T_{\rm N}^{\rm RG}$.
We see that  $l_{\rm cross}>l_N$ (i.e. $T_{\rm cross}<T_{\rm N}^{\rm RG}$) for $ t_{b 0}=0.02E_0$,
 while  $l_{\rm cross}<l_N$ 
(i.e. $T_{\rm cross}>T_{\rm N}^{\rm RG}$) for $ t_{b; 0}=0.3E_0$.
As is seen from Fig.~7(b),  $t_{b0}=t_{b}^\ast=0.108E_0$ gives the   critical value where $l_{\rm cross}=l_N$ (i.e. $T_{\rm cross}=T_{\rm N}^{\rm RG}$).
In Fig.~8,  dependence of $t_{b}^{\ast}$  
  on the dimerization ratio $\delta$ is shown for $U/\pi v_F=1.0,1.45,1.6$.  
We see that a finite $\delta$ causes a finite $t_{b}^{\ast}$. This situation arises, since 
$\delta$ strengthens   the  intrachain umklapp scattering [see eq.(\ref{eqn:Uvsg})],    and consequently
 more severely suppresses 
the interchain one-particle propagation. As $U$ increases, overall magnitude of $t_{b}^{\ast}$ increases,
 since the increasing $U$ strengthens both the umklapp amd mormal scattering 
and consequently enhances the ICEX mechanism.

\section{Nesting-driven SDW phase Transition in the Fermi Liquid Regime}

As discussed in the previous section, for  $t_{b}^{\ast}<t_{b0}$, 
the system undergoes a crossover to  the FL regime below $T_{\rm cross}$. 
Inside the FL regime,  
the  SDW phase transition is driven by the Fermi surface nesting. 
Except the region  with $t_{b0}$
 very near the critical value, $t_{b}^\ast$,  the nesting-driven SDW phase transition can be treated by the random-phase-approximation (RPA) where
only   unrenormalized particle-hole fluctuations are taken into account.\cite{Yamaji82,HF86,Shimahara88} 
Quite recently, Kino and Kontani treated the effects of the particle-hole fluctuations on the
one-particle propagator in a consistent manner.\cite{KinoKontani99}
Near the critical value,   interference between the particle-particle and particle-hole polarization is expected to be so strong that the
RPA  treatment   for the phase transition would  be insufficient. 
In this work,  to give a qualitative view on  the different nature of the SDW transitions from the incoherent metal regime and from the FL regime,
 we simply treat the phase transition in the FL regime using the
RPA. 

In Fig.~9, we show a series of diagrams representing the particle-hole fluctuations which contribute to the 
 transverse susceptibility.
The longitudinal counterpart is given similarly.
We here stress that, in Fig.~9, the unrenormalized one-particle propagator
has a two-dimensional character  as
\be
{\cal G}_{\nu}^{\rm 2D} (\k,\varepsilon)=[ \ii \varepsilon -v_F(\mid\!k\!\mid-k_F)+2t_b \cos k_b-2t_{b}'\cos 2 k_b]^{-1},
\en
where  $\nu=R$ or $L$ and $\varepsilon$ is a fermion thermal frequency.
The  SDW phase transition temperature, $T^{\rm RPA}_{N}$, 
  is determined through the condition for 
the dimensionless static  particle-hole polarization function at an optimal nesting vector, $\chi(T;t_{b0})$,
\be
 \chi(T^{\rm RPA}_{\rm N};t_{b0})&=&[g_{2;0}+g_{3;0}]^{-1}\non\\
&=&\pi v_F\left[{U\over 2}\left(1+{2\delta\over 1+\delta^2}\right) \right]^{-1}.\label{eqn:RPAcond}
\en
The polarization function in the noninteracting case, $\chi(T;t_{b0})$, is given  by
\begin{eqnarray}
\chi(T;t_{b0})&=&-{\pi v_F }T\sum_{\stackrel{\mid  \xi_R(k_a)\mid \leq E_0/ 2}{\mid k_b\mid\leq \pi}}
\sum_{\varepsilon}
{\cal G}_{R}^{\rm 2D}(\k,\varepsilon){\cal G}_{L}^{\rm 2D}(\k-\Q,\varepsilon)\non\\
&=&{1\over 4\pi }\int_{-E_0/2}^{E_0/2}d\xi_{R}\int_{0}^{\pi}{dk_y}
{\tanh  [\xi(\k)/ 2 T  ]
-\tanh  [\xi(\k-\Q)/ 2 T]
\over   \xi(\k)  -\xi(\k-\Q)
},\label{eqn:sus}
\end{eqnarray}
where we  adopt the intrachain linearized bandwidth cutoff, $E_0$, to keep consistency with the treatment in the 
 previous section. 

In Fig.10, we show dependence of $T^{\rm RPA}_{\rm N}$ on $t_{b0}$ for $\delta=0.05,0.2,0.4$ in the case
of the commensurate SDW vector,
 \be
\Q=(\pi,\pi).
\en
We here put $W=8E_0$.
$T^{\rm RPA}_{\rm N}$ is sharply suppressed at the critical vales, $t_{b0:\rm cr}/E_0\sim 0.30,0.347,0.39$, for  
  dimerization ratios, $\delta=0.05,0.2,0.4$,
respectively.
Increasing $t_{b0}$ weakens the degree of nesting  through the third term of (\ref{eqn:1Pdispersion})   and
consequently reduces the transition temperature.
As $\delta$ decreases, overall magnitude of $T_{\rm N}^{\rm RPA}$ decreases, since the decreasing $\delta$ weakens the umklapp scattering strength
and consequently reduces the effective two-particle interaction strength [see eq.~(\ref{eqn:RPAcond})].
As pointed out by Hasegawa and Fukuyama,\cite{HF86} there exists  a very narrow region of $t_{b0}$ in the vicinity of
$t_{b0:\rm cr}$ where  an incommensurate SDW  is favored, but we do not go into the minor details on this issue here.

\section{Phase Diagrams}

By combining the results on $T_{\rm N}^{\rm RG}$, $T_{\rm cross}$ 
and $T_{\rm N}^{\rm RPA}$ in \S 3 and 
\S 4,
we obtain  a phase diagram   covering
both the incoherent metal regime and the FL   regime.
In Fig.~11, we show a phase diagram   for  $\delta=0.2$, $U=1.45\pi v_F$. 
In this case, $T_{\rm N}^{\rm RG}$ and $T_{\rm N}^{\rm RPA}$ meet together
around the critical value,  $t_{b0}=t_{b}^\ast$.
This choice of parameters is   consistent with that of the TMTTF compound.\cite{Mila95}
As shown in Fig.~8, $t_{b}^\ast$ is sensitive to both $U$ and $\delta$.
Thus, by tuning   $U$ and $\delta$  appropriately,  a  phase diagram qualitatively
similar to  Fig.~11 is always obtained, as far as
the low-energy asymptotics of the system without $t_{b0}$ is a Mott insulator.
The $t_{b0}$-dependence of the SDW phase transition temperature in the two regimes is interpreted as follows.
For $t_{b0}<t_{b}^\ast$, the increasing $t_{b0}$ enhances the interchain  propagation of the dominant 1DAF correlation
through the ICEX process toward low temperatures and consequently increases the SDW transition temperature.
This situation is consistent with Bourbonnais' argument based on the preformed Mott gap, $\Delta_{\rho}$,\cite{Bourbonnais97} that
 the Stoner criteria gives the N$\acute{\rm e}$el temperature, $T_{\rm N}\sim {\tilde t_b}^2/\Delta_{\rho}$, 
which increases when the renormalized interchain one-particle hopping integral, $\tilde t_b$, increases    
under pressure.
On the other hand,
once the system undergoes the crossover to the FL regime for $t_{b0}>t_{b}^\ast$,
the increasing  $t_{b0}$   weakens  the degree of nesting of the Fermi surface  and
consequently decreases the SDW transition temperature.

When we adopt    dimerization ratios smaller  or larger than $\delta=0.2$
 with the same interaction strength
as in Fig.~11,
 both  $T_{\rm N}$ and $T_{\rm N}^{\rm RPA}$ change accordingly  and  no longer meet together 
 around  the  critical value,  $t_{b0}=t_{b}^\ast$.  
As shown in Fig.~12, 
$T_{\rm N}^{\rm RPA}$ excessively dominates $T_{\rm N}$ for  $\delta=0.05$~[Fig.~12(a)],
while the opposite situation occurs for $\delta=0.6$~[Fig.~12(b)].
In the case of $\delta=0.05$, the intrachain umklapp scattering is much weaker than that in the case of $\delta=0.2$
and consequently  the ICEX mechanism becomes ineffective over the wide range of $t_{b0}$.
On the other hand, in the case of $\delta=0.6$, the strong intrachain umklapp scattering enhances the ICEX mechanism.
These   misfits should be seen as an artifact originating from the following two facts.
First, for $t_{b0}<t_{b}^\ast$, feedback effects of the interchain processes on the intrachain processes in the 
1D regime have not taken into account
within the present scheme.
Secondly, for $t_{b0}>t_{b}^\ast$, we have treated the SDW phase transition in the simple RPA and  
have not taken account of
the effects of strong quantum fluctuations in the vicinity of $t_{b}^\ast$.
Concerning these points,
we   did not apply the renormalization-group (RG) procedure to the $g_{2;l}$ and $g_{3;l}$ processes 
in (\ref{eqn:RPAcond}).
However, in the vicinity of $t_{b}^\ast$, the RG treatment of $g_{2;l}$ and $g_{3;l}$ 
might still work.\cite{Bourbonnais97} 
If  these  quantum-fluctuation effects
are fully taken into account over the whole range of $t_{b0}$, the resultant SDW phase transition temperature
 is expected to exhibit
a continuous change  similar to Fig.~11.
In any case, Fig.~12 suggests that the ICEX mechanism is more sensitive to the dimerization ratio, $\delta$,
than the nesting mechanism.

Although our model considered here misses some details of real structure of the TMTTF and TMTSF compounds such as misfit 
between the adjacent chains, which is discussed
  in \S 5.2,
our results are qualitatively consistent with the experimental suggestions that 
the metallic phases just above $T_N$ in the TMTTF and TMTSF compounds
at ambient pressure
belong to the different regimes where 
the 2D (or 3D) quasiparticle coherence is present  in (TMTSF)$_2$PF$_6$, but
absent in (TMTTF)$_2$Br.
As shown in Fig.~8,  the  critical value, $t_{b}^{\ast}$, is sensitive to the dimerization ratio, $\delta$, since $\delta$ controls the
intrachain 
umklapp scattering strength. This result suggests
that, in the  TMTTF compounds with larger dimerization ratio,   the ICEX-driven SDW transition  
becomes effective  up to  a considerable magnitude of $t_{b0}$, while  
in the case of TMTSF with smaller dimerization ratio, the ICEX mechanism becomes ineffective  even for rather small
$t_{b0}$.
This situation suggests that  very small $t_{b0}$ is sufficient for
(TMTSF)$_2$PF$_6$ to evolve  the interchain coherent one-particle motion  and is consistent with the experimental suggestion
and   phenomenological discussion  by Emery {et al.}\cite{Emery82} 

Roughly speaking,    the increasing $t_{b0}$ corresponds to   increasing applied pressure.
Thus our SDW phase diagram,     Fig.~11, is  consistent with the
experimental observation in  
 (TMTTF)$_2$Br under pressure, $P$, where  
the transition temperature increases  at   $P<P_{\rm opt}=5$kbar, while decreases    
  at  $P>P_{\rm opt}$.\cite{Klemme96}

\section{Discussion}

 In this section,  we  
  discuss relevance of the present work
to other ones and comment on the problems which remain unsettled in the present scheme.

\subsection{Zero-temperature Mott gap}

In the present paper, we have assumed the scaling hypothesis works well down to $T_{\rm N}^{\rm RG}$
for $t_{b0}<t_{b}^\ast$
and determined $T_{\rm N}^{\rm RG}$   based solely on  
the 2-loop renormalization-group flows without introducing the preformed Mott gap.
As stressed repeatedly,  
in our formulation, growth of the umklapp scattering toward low temperatures plays an essential role on  occurrence of
the SDW phase transition for $t_{b0}<t_{b}^\ast$. The growth of the umklapp scattering strength is regarded as  
 preformation of the Mott gap.
Thus
our results are consistent with the views based on the Mott gap.\cite{BourbonnaisCaron91,Bourbonnais93a,Bourbonnais93b,Bourbonnais97,
Giamarchi97,Suzumura98,Tsuchiizu99a,Tsuchiizu99b}

To see this situation more closely,  
we here discuss  the zero-temperature Mott gap of the isolated dimerized quarter-filled 
Hubbard chain in the weak-coupling regime:\cite{PencMila94,Tsuchiizu99c}
\begin{eqnarray}
\Delta_{\rho0}(U, \delta)={4v_F\over  \pi} 
 (1-A^2)^{1/4}\exp\left[-{1\over4}{\tanh^{-1}A\over A}+{1\over 4}+{\tilde C(A)}\right] \sqrt{ U\over v_F}\exp\left[-{2\pi v_F \over U}{\tanh^{-1}A\over A}\right], \label{eqn:MottGap}
\end{eqnarray}
where the Fermi velocity of the noninteracting chain is given
 by $v_F={\sqrt{2}\over 8}W(1-\delta^2)/\sqrt{1+\delta^2}$, $A=(1-\delta^2)/(1+\delta^2)$ and $\tilde C(A)$ is a function of   $A$.\cite{seeref}
We note that $v_F$ depends on the total bandwidth, $W$, while the temperature scale in the RG
scheme is given in terms of the linearized bandwidth, $E_0$.
 Thus, we should bear in mind that there is arbitrariness in specification of
 the quantitative temperature scale corresponding to $\Delta_{\rho0}$. 
In the case of  $W=8E_0$ and $U=2.5 t_{a1}$, we estimate and show  
$\delta$-dependence of $\Delta_{\rho0}$, $t_{b}^\ast$ and $T_{\rm um}$, in the unit of $t_{a1}$,   in Fig.~13.
$T_{\rm um}$ is defined as the temperature scale at which the intrachain umklapp scattering strength, $g_{3;l}$, reaches unity.
We see that two energy scales, $\Delta_{\rho0}$ and $T_{\rm um}$, are close to each other.
 This fact indicates that  
 growth of the umklapp scattering strength qualitatively corresponds to the preformation of the Mott gap.

We also comment on relevance of the present results to the
confinement-deconfinement transition  in terms of  $\Delta_{\rho 0}$
 and  $t_{b0}$.\cite{Suzumura98,Tsuchiizu99a,Tsuchiizu99b}
As is seen from Fig.~13,   $t_{b}^\ast > \Delta_{\rho 0}$ for smaller dimerization, $\delta\leq 0.22$. 
According to arguments given by Suzumura and Tsuchiizu\cite{Suzumura98,Tsuchiizu99a,Tsuchiizu99b}
for the   coupled two half-filled chains,
this fact might indicate that   the incoherent metal regime for $t_{b0}\stackrel{<}{\sim}t_{b}^\ast$ 
in the phase diagram  of Fig.~11 belongs to the deconfinement regime.
However, it is beyond the scope of this paper to reconcile  the 2-loop RG flow of   $t_{b;l}$ determined by eqs.~(\ref{eqn:rgeqt1})
and (\ref{eqn:rgeqt2})
with a possibility of the confinement in the present case of the infinite number of chains.

\subsection{Misfit between the adjacent chains}

 We have not treated the misfit between the adjacent chains existing  in
the actual TMTTF and TMTSF crystals, which causes  two kinds of interchain  one-particle hopping integrals.\cite{Grant83}
In the FL regime, $t_{b0}>t_{b}^\ast$, it is straightforward to take account of the   misfit   by starting with the corresponding
one-particle dispersion.\cite{Yamaji82} 
On the other hand, serious treatment of the misfit effects in the incoherent metal regime
is beyond the scope of the RG analysis, since we take a continuum limit along the $a$-axis in the RG analysis.
It would be essential    to take account of  the misfit effects in both the two regimes
  to clarify the reason why
the experimentally suggested SDW vector  of   (TMTSF)$_2$PF$_6$
 is $(1/2,0.24\pm 0.03,-0.06\pm 0.20)$,\cite{Takahashi86}
 while that of  (TMTTF)$_2$Br
is $(1/2,1/4,0)$,\cite{Nakamura95} in the unit of the reciprocal lattice constants, $a^\ast$, $b^\ast$ and $c^\ast$.

\subsection{Effects of the nearest-neighbor  Coulomb repulsion}

We  comment on the effects of the intrachain nearest-neighbor  Coulomb repulsion,  $V$.
In the present paper,
we have not considered $V$,  since the
  nature of the SDW transition approached {\it from the metallic side} 
 is insensitive to $V$ in the following reason. 
In the 1D regime, as far as the low-energy asymptotics of the   system without $t_{b0}$ is a Mott insulator,
 the presence of $V$ only   modifies the unrenormalized values of the scattering strengths, (\ref{eqn:Uvsg}). 
In the FL regime,  the  SDW phase transition determined through the RPA  condition, (\ref{eqn:RPAcond}),
is insensitive to $V$, unless $V$ is  too large to destabilize the SDW phase transition.
We note  that the effects of $V$ and the next-nearest-neighbor repulsion, $V_2$,
  become  important, if we clarify
the {\it  real space structure } of the spin and charge  ordering  {\it inside} the SDW phase.\cite{SeoFukuyama97,Kobayashi98}

\section{Summary}

In this work, we have studied the SDW phase transitions in the dimerized quarter-filled chains
weakly coupled via the interchain one-particle  hopping, $t_{b0}$.
It is  shown that there exists a  critical value of $t_{b0}$, $t_{b}^\ast$, between the
incoherent metal regime ($t_{b0}<t_{b}^\ast$) and the Fermi liquid regime ($t_{b0}>t_{b}^\ast$) in the metallic phases
above the SDW transition temperature.
For $t_{b0}<t_{b}^\ast$, we assumed that the scaling hypothesis in the 1D regime  works well down to the
transition temperature, $T_{\rm N}^{\rm RG}$, and discussed the ICEX-driven SDW phase transition 
from the incoherent metal phase, based solely on the 2-loop RG flows.
 In our formulation, growth of the umklapp scattering  toward low temperatures, which is regarded as the preformation of
the 1D Mott gap,  plays an essential role on  occurrence of
the SDW phase transition from the incoherent metal phase.

On the other hand, for $t_{b0}>t_{b}^\ast$, 
 the system undergoes the crossover to the FL regime around the   temperature, $T_{\rm cross}$.
In this case, 
the increasing  $t_{b0}$   weakens  the degree of nesting of the Fermi surface  and
consequently  decreases the nesting-driven SDW phase transition temperature.

\acknowledgements

We acknowledge  fruitful discussions with  Y. Suzumura, M. Tsuchiizu, H. Kino,   H. Kontani and
T. Nakamura. 
This work was supported by a Grant-in-Aid for Encouragement of Young Scientists from the Ministry of Education, Science, Sports and Culture, Japan.

\appendix

\section{Derivation of the Renormalization-Group Equations}
The renormalization-group procedure  is best formulated in the path-integral
 representation of the partition function,\cite{BourbonnaisCaron91}
\be
Z=\int {\cal D} e^{S},
\en
where $S$ is the action of the system and $\cal D$ symbolizes the measure of the path-integral over the fermionic Grassmann
variables.
The action at the energy scale specified by $l$ contains the four  parts corresponding to the equation, (\ref{eqn:Hamiltonian}),
\be
S=S_{a;l}^{(1)}+S_{a;l}^{(2)}+S_{b;l}^{(1)}+S_{b;l}^{(2)},
\en
which are give by
\begin{eqnarray}
S_{a;l}^{(1)}&=&\sum_{k_{a}\in {\cal C}_{l}} \sum_{-\pi\leq k_b\leq \pi}\sum_{\varepsilon} 
\sum_{\sigma}\left[{\cal G}_{R}^{-1}(K_a)
R^{\ast}_{\sigma}(K)R_{\sigma}(K)
+{\cal G}_{L}^{-1}(K_{a})
L^{\ast}_{\sigma}(K)L_{\sigma}(K)\right],\\
S_{b;l}^{(1)}&=&2\sum_{k_a\in {\cal C}_{l }}\sum_{-\pi\leq k_{b}\leq \pi}\sum_{\varepsilon}\sum_{\sigma}
(t_{b;l}\cos   k_b -t_{b;l}'\cos  2 k_b)
\left[L^{\ast}_{\sigma}(K)L_{\sigma}(K) 
+ R^{\ast}_{\sigma}(K)R_{\sigma}(K)\right],\\
S_{a;l}^{(2)}&=&
-{\pi v_{F} T\over N_a N_b }  \sum_{k_{ai}\in C_{l }}\sum_{-\pi\leq k_{bi}\leq \pi}\sum_{\varepsilon_i,\sigma_i}
 g_{l}^{\sigma_1\sigma_2\sigma_3\sigma_4} 
R^{\ast}_{\sigma_1}(K_{1})L^{\ast}_{\sigma_2}(K_{ 2}) 
L_{\sigma_3}(K_{3}) R_{\sigma_4}(K_{ 4}) \\
&-&{\pi v_{F}T\over 2N_a N_b  }\sum_{k_{ai}\in C_{l }}\sum_{-\pi\leq k_{bi}\leq \pi}\sum_{\varepsilon_i,\sigma,\sigma'}  g_{3;l} 
% \\&&
\left[
R^{\ast}_{\sigma}(K_{1})R^{\ast}_{ \sigma'}(K_{2}) 
L_{\sigma' }(K_{3}) L_{\sigma }(K_{4})+{\rm c.c}\right],\non\\
S_{b;l}^{(2)}&=& -{\pi v_{F}T\over   N_a N_b} \sum_{Q} 
  J_{ q_b ;l} 
{\SDW}_{l }(\qpara,q_b,\omega )\cdot{\SDW}_{l }^\ast(\qpara,q_b, \omega ) \\
&-&{\pi v_{F}T\over   N_a N_b}  \sum_{Q}  K_{ q_b ;l}  
\left[
{\SDW}_{l } (2k_F+\qpara, q_b,\omega )\cdot{\SDW}_{l }(2k_F-\qpara,-q_b, -\omega )
+{\rm c.c}\right], \non
\end{eqnarray}
where
$R_{\sigma}$ and $L_{\sigma}$ are  Grassman variables representing the right- and left-moving electrons, respectively,
$K_a=(k_a,\varepsilon)$ and $K=(k_a,k_b,\varepsilon)$   with $\varepsilon$ being a fermion thermal frequency.
The intrachain one-particle propagator is given by
\be
{\cal G}_\nu (K_a)=[ \ii \varepsilon -\xi_{\nu}(k_a)]^{-1}.
\en

We split up the set of $k_a$-points, ${\cal C}_{l}$, into  
two subsets  as ${\cal C}_{l}={\cal C}_{l+\dd l}^{<}\oplus d{\cal C}_{l+\dd l}^{>}$, where
$
{\cal C}_{l+\dd l}^{<}\equiv\left\{k_a \mid \mid \xi_\nu(k_a)\mid\leq \omega_{l+\dd l}/2 \right\}
$
and
$
d{\cal C}_{\nu;l+\dd l}^{>}\equiv\left\{k_a  \mid \omega_{l+\dd l} /2\leq \mid 
\xi_\nu(k_a)\mid\leq \omega_{l}/2 \right\},
$
represent the low- and high-energy shells, respectively.
Accordingly,   the     action is decomposed as
$
S_l=S^{<}_{l+\dd l}+S^{>}_{l+\dd l}.
$
Integration over the modes in the high-energy shell  gives  
\begin{eqnarray}
Z=\ds\int_{ {\cal C}_{l+\dd l}^{<}}
{\cal D} \exp\left[S_{l+\dd l}^{<}+\sum_{p,q,r=1}^{\infty}
\Gamma_{pqr}\right] ,\label{eqn:avefast}
\end{eqnarray}
where $\ds\int_{ {\cal C}_{l+\dd l}^{<}}{\cal D}$ means  that the Fermion momenta   are restricted to the low-energy shell.
All the renormalization effects come from the perturbative expansion
\be
\Gamma_{pqr}={1\over p!q!r!}\langle\!\langle 
[S_{a;l+\dd l}^{(2)>}]^{p}[S_{b;l+\dd l}^{(1)>}]^{q}[S_{b;l+\dd l}^{(2)>}]^{r}\rangle\!\rangle_{\rm c}.
\en
The average over the modes in the high-energy shell  is defined as
$
\langle\!\langle(\cdots)
\rangle\!\rangle\ds
=Z_{>}^{-1}\int_{ d{\cal C}_{l+\dd l}^>}{\cal D}  \exp[S_{a;l+\dd l}^{(1)>}]\,\,
(\cdots), 
$
with 
$
Z_{>}=\ds\int_{ d{\cal C}_{l+\dd l}^>} {\cal D}  \exp[S_{a;l+\dd l}^{(1)>}]
$
and the subscript 'c' represents the connected diagrams.
We perform a perturbative expansion 
by picking up the  Feynmann 
diagrams whose contribution is in proportion to $\dd l$ and then replacing
$S_{l+\dd l}^{<}+\sum_{p,q,r=1}^{\infty}
\Gamma_{p,q,r}$  with
the renormalized action.

Then the renormalized action  is  
written in the form,
\begin{eqnarray}
\tilde S^{<}_{l+\dd l}&=&\sum_{k_{a}\in {\cal C}_{l+\dd l}} \sum_{-\pi\leq k_b\leq \pi}\sum_{\varepsilon} 
\sum_{\sigma}[1+\theta_l \dd l]\non\\
&&\left[{\cal G}_{R}^{-1}(K_a)
R^{\ast}_{\sigma}(K)R_{\sigma}(K)
+{\cal G}_{L}^{-1}(K_{a})
L^{\ast}_{\sigma}(K)L_{\sigma}(K)\right]\non\\
&+&2\sum_{k_a\in {\cal C}_{l+\dd l}}\sum_{-\pi\leq k_{b}\leq \pi}\sum_{\varepsilon}\sum_{\sigma}
(t_{b;l}\cos   k_b -t_{b;l}'\cos  2 k_b)
\left[L^{\ast}_{\sigma}(K)L_{\sigma}(K) 
+ R^{\ast}_{\sigma}(K)R_{\sigma}(K)\right]\non\\
&-&{\pi v_{F} \over N_a N_b }  \sum_{k_{ai}\in C_{l+\dd l}}\sum_{-\pi\leq k_{bi}\leq \pi}\sum_{\varepsilon_i,\sigma_i}
\left[g_{l}^{\sigma_1\sigma_2\sigma_3\sigma_4}+w_{l}^{\sigma_1\sigma_2\sigma_3\sigma_4}\dd l\right]
R^{\ast}_{\sigma_1}(K_{1})L^{\ast}_{\sigma_2}(K_{ 2}) 
L_{\sigma_3}(K_{3}) R_{\sigma_4}(K_{ 4})  \non\\
&-&{\pi v_{F}\over 2N_a N_b  }\sum_{k_{ai}\in C_{l+\dd l}}\sum_{-\pi\leq k_{bi}\leq \pi}\sum_{\varepsilon_i,\sigma,\sigma'} \left[g_{3;l}+w_{3;l}\dd l\right]
%\non\\&&
\left[
R^{\ast}_{\sigma}(K_{1})R^{\ast}_{ \sigma'}(K_{2}) 
L_{\sigma' }(K_{3}) L_{\sigma }(K_{4})+{\rm c.c}\right]\non\\
 &-&{\pi v_{F}\over  N_a N_b} \sum_{Q} 
\left[f^{J}_{ q_b ;l}\dd l+w^{J}_{ q_b ;l} J_{ q_b ;l}\dd l \right]
{\SDW}_{l+\dd l}(\qpara,q_b,\omega )\cdot{\SDW}_{l+\dd l}^\ast(\qpara,q_b, \omega ) \\
&-&{\pi v_{F}\over  N_a N_b}  \sum_{Q} 
\left[f^{K}_{ q_b ;l}\dd l+w^{K}_{ q_b ;l} K_{ q_b ;l}\dd l \right]
\left[
{\SDW}_{l+\dd l} (2k_F+\qpara, q_b,\omega )\cdot{\SDW}_{l+\dd l}(2k_F-\qpara,-q_b, -\omega )
+{\rm c.c}\right]\non,
\end{eqnarray}
with
\be
w_{l}^{\sigma_1\sigma_2\sigma_3\sigma_4}=\delta_{\sigma_1\sigma_4}\delta_{\sigma_2\sigma_3}w_{2;l}
-\delta_{\sigma_1\sigma_3}\delta_{\sigma_2\sigma_4}w_{1;l}.
\en

Next, to restore the original cutoff,  we   rescale  the momenta and  frequencies as  
$
\tilde  K_a={\rm e}^{\dd l}K_a.
$
Here we note that the wave number in the $b$-axis direction is {\it not} rescaled.  
We   perform the field-renormalization   as
\be
\tilde  R_{\sigma}(\tilde K)=[1+{1\over 2}(\theta_l -3) \dd l]  R_{\sigma}(K), \label{eqn:fieldrenormalization}
\en
with
$\tilde K=( \tilde k_{a},k_{b}, \tilde\varepsilon)$,
and rewrite $\tilde S_{l+\dd l}^{<}$ as
 \begin{eqnarray}
&&\tilde S^{<}_{l+\dd l}=\sum_{\tilde k_{a}\in {\cal C}_{l}} \sum_{k_b,\tilde \varepsilon,\sigma} \left[{\cal G}_{R}^{-1}(\tilde K_a)
\tilde R^{\ast}_{\sigma}(\tilde K)\tilde R_{\sigma}(\tilde K)
+{\cal G}_{L}^{-1}(\tilde K_{a})
\tilde L^{\ast}_{\sigma}(\tilde K)\tilde L_{\sigma}(\tilde K)\right]\non\\
&+&2\sum_{\tilde k_a\in {\cal C}_{l}}\sum_{k_{b},\tilde \varepsilon,\sigma}(1-\theta_l \dd l)(t_{b;l}\cos   k_b 
-t_{b;l}'\cos  2 k_b)
\left[\tilde L^{\ast}_{\sigma}(\tilde K)\tilde L_{\sigma}(\tilde K) 
+ \tilde R^{\ast}_{\sigma}(\tilde K)\tilde R_{\sigma}(\tilde K) \right]\non\\
&-&{\pi v_{F} \over N_a N_b }  \sum_{\tilde k_{ai}\in C_{l}}\sum_{k_{bi},\tilde \varepsilon_i,\sigma_i}
\left[g_{l}^{\sigma_1\sigma_2\sigma_3\sigma_4}+(w_{l}^{\sigma_1\sigma_2\sigma_3\sigma_4}-2\theta_lg_{l}^{\sigma_1\sigma_2\sigma_3\sigma_4})\dd l\right] 
\tilde R^{\ast}_{\sigma_1}(\tilde K_{1})\tilde L^{\ast}_{\sigma_2}(\tilde K_{ 2}) 
\tilde L_{\sigma_3}(\tilde K_{3}) \tilde R_{\sigma_4}(\tilde K_{ 4})  \non\\
&-&{\pi v_{F}\over 2N_a N_b  }\sum_{\tilde k_{ai}\in C_{l}}\sum_{k_{bi},\tilde \varepsilon_i,\sigma,\sigma'} \left[g_{3;l}+(w_{3;l}-2\theta_l g_{3;l}) \dd l\right]
%\non\\&&
\left[
\tilde R^{\ast}_{\sigma}(\tilde K_{1})\tilde R^{\ast}_{ \sigma'}(\tilde K_{2}) 
\tilde L_{\sigma' }(\tilde K_{3}) \tilde L_{\sigma }(\tilde K_{4})+{\rm c.c}\right]\non\\
 &-&{\pi v_{F}\over  N_a N_b} \sum_{\tilde Q} 
\left[J_{ q_b ;l}+(w^{J}_{ q_b ;l} -2\theta_l J_{ q_b ;l})  \dd l \right]
\tilde {\SDW}_{l}(\tilde \qpara,q_b,\tilde \omega )\cdot\tilde {\SDW}_{l}^\ast(\tilde \qpara,q_b, \tilde\omega ) \\
&-&{\pi v_{F}\over  N_a N_b}  \sum_{\tilde Q} 
\left[K_{ q_b ;l}+(w^{K}_{ q_b ;l} -2\theta_l K_{ q_b ;l})\dd l \right]
\left[
\tilde {\SDW}_{l} (2\tilde k_F+\tilde \qpara, q_b,\tilde\omega )\cdot\tilde {\SDW}_{l}(2\tilde k_F-\tilde \qpara,-q_b, -\tilde\omega )
+{\rm c.c}\right]\non.
\end{eqnarray}

By identifying the renormalized quantities with the quantities at the energy scale specified by $l+\dd l$, we obtain
\be
g_{l+\dd l}^{\sigma_1\sigma_2\sigma_3\sigma_4}&=&
g_{l}^{\sigma_1\sigma_2\sigma_3\sigma_4}+(w_{l}^{\sigma_1\sigma_2\sigma_3\sigma_4}-2\theta_lg_{l}^{\sigma_1\sigma_2\sigma_3\sigma_4})\dd l, \\
g_{3;l+\dd l}&=&g_{3;l}+(w_{3;l}-2\theta_l g_{3;l}) \dd l, \\
t_{b;l+\dd l}&=&t_{b;l}(1-\theta_l \dd l), \\
t_{b;l+\dd l}'&=&t_{b;l}'(1-\theta_l \dd l),\\
J_{ q_b ;l+\dd l}&=&J_{ q_b ;l}+(w^{J}_{ q_b ;l} -2\theta_l J_{ q_b ;l})  \dd l,\label{eqn:Jqbr}\\
K_{ q_b ;l+\dd l}&=&K_{ q_b ;l}+(w^{K}_{ q_b ;l} -2\theta_l K_{ q_b ;l})  \dd l\label{eqn:Kqbr}.
\en

In Figs.~4 and 6,  we show   the  diagrams which contribute to $\theta_l$~[Fig.~4], 
$w^{J}_{ q_b ;l}$~[Fig.~6(a)]
and $w^{K}_{ q_b ;l}$~[Fig.~6(b)], respectively.
In  the intrachain self-energy corrections ($\theta_l$) and 
the intrachain 
vertex corrections ($w_{l}^{\sigma_1\sigma_2\sigma_3\sigma_4}$ and $w_{3;l}$),
we neglect  the contributions from the interchain   interaction.  
It should  be also noted that there are no    2-loop contributions  to  $w^{J}_{ q_b ;l}$ 
and $w^{K}_{ q_b ;l}$ 
 because of the constraint on the chain indices for the 3rd order diagrams.
After lengthy but straightforward manipulation,
we obtain the following expressions,

\be
\theta_l&=&{1\over 4} \left[g_{1;l}^2+g_{2;l}^2-g_{1;l}g_{2;l}+ g_{3;l}^2/2\right],\\
w_{1;l}&=&
-g_{1;l}^2
+{1\over 2}   g_{1;l} g_{2;l}^2
-{1\over 2} g_{1;l}^2  g_{2;l} 
+{1\over 4} g_{1;l} g_{3;l}^2,\\
w_{2;l}&=&
-{1\over 2} g_{1;l}^2
+{1\over 2}  g_{3;l}^2
+{1\over 2} g_{2;l}^3
+{1\over 2}  g_{1;l}^2 g_{2;l}
-{1\over 2} g_{1;l}  g_{2;l}^2\non\\
& &-{1\over 4} g_{1;l}^3
+{1\over 4} g_{1;l}  g_{3;l}^2
-{1\over 4} g_{2;l}  g_{3;l}^2,\\
w_{3;l}&=&
- g_{1;l}g_{3;l}
+2 g_{2;l} g_{3;l}
-{1\over 2} g_{2;l}^2 g_{3;l}
+{1\over 2}  g_{1;l}   g_{2;l} g_{3;l}
+{1\over 4}  g_{1;l}^2 g_{3;l},\\
w^{J}_{ q_b ;l}&=&{1\over 2} \left[{g_{2;l}} ^2+4{g_{3;l}}^2\right]f_{l}(q_b)
\non\\
&+&{1\over 2}\left[g_{2;l} J_{q_b;l} +4g_{3;l}  K_{q_b;l }\right]  
-{1\over 4}\left[J_{q_b;l}^2+4K_{q_b;l}^2\right],\label{eqn:WJ}\\
w^{K}_{ q_b ;l}&=&2 
{g_{2;l}}{g_{3;l}}f_{l}(q_b)
+2\left[g_{2;l} K_{q_b;l} +g_{3;l} J_{q_b;l} \right]-J_{q_b;l}K_{q_b;l},\label{eqn:WK}
\en
which give the RG equations (\ref{eqn:rgeqt1}), (\ref{eqn:rgeqt2}),
(\ref{eqn:rgeqg1}) $\sim$ (\ref{eqn:rgeqg3}), (\ref{eqn:rgeqJ}) and (\ref{eqn:rgeqK}).
The terms including $\theta_l$ in (\ref{eqn:Jqbr}) and (\ref{eqn:Kqbr}) come from the  field-renormalization due to the intrachain self-energy processes
and give  negligibly small corrections to  the solutions.

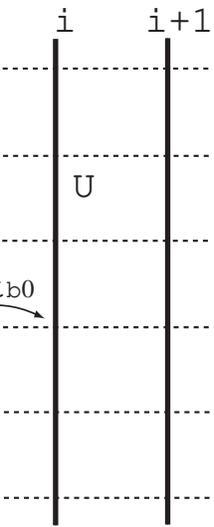
\begin{figure}
\caption{ Array of  
 dimerized quarter-filled Hubbard chains considered here.}
\end{figure}

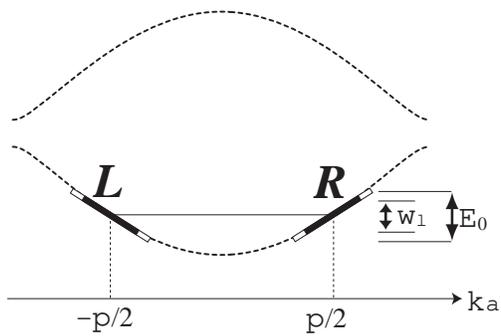
\begin{figure}
\caption{Broken lines represent the one-particle dispersion, (\ref{eqn:1Pdis}), in the absence of $t_{b0}$ and $U$.
$R$ and $L$ are the linearized  dispersions for the right-  and left-moving electrons with the bandwidth cutoff $E_0$.
At the the energy scale, 
$
\omega_l=E_{0}{\rm e}^{-l},
$   we observe the system.
}
\end{figure}

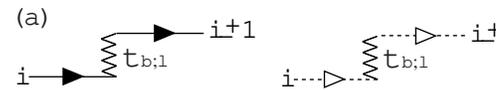
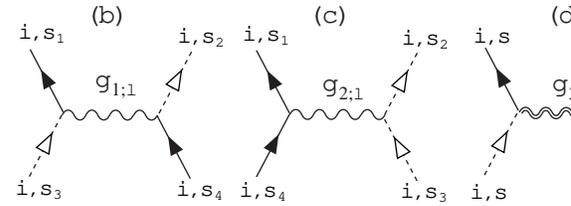
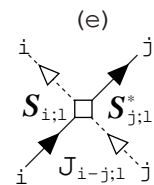
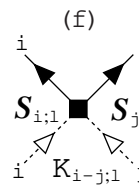
\begin{figure}
\caption{ Fundamental processes considered here. The 
solid and broken lines represent the propagators for the right-  and 
left-moving electrons, respectively.
The   zigzag line in (a) represents  the  
 interchain one-particle hopping process.
Single and double wavy lines  in (b), (c) and (d)
represent  the   intrachain   {\lq\lq}normal{\rq\rq}   and {\lq\lq}umklapp{\rq\rq}  scattering, respectively. 
White and black squares in (e) and (f) 
 represent  the  interchain    interaction between the $i$-th and $j$-th chains in
the    antiferromagnetic channel, due to the normal  (e) and
  umklapp (f) processes.
 }
\end{figure}

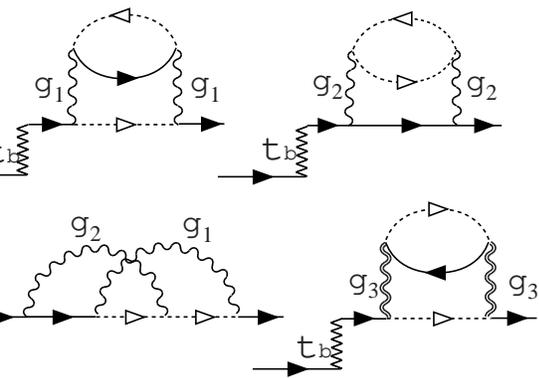
\begin{figure}
\caption{Renormalization   of the interchain  one-particle propagation through
 the self-energy processes.
}\end{figure}

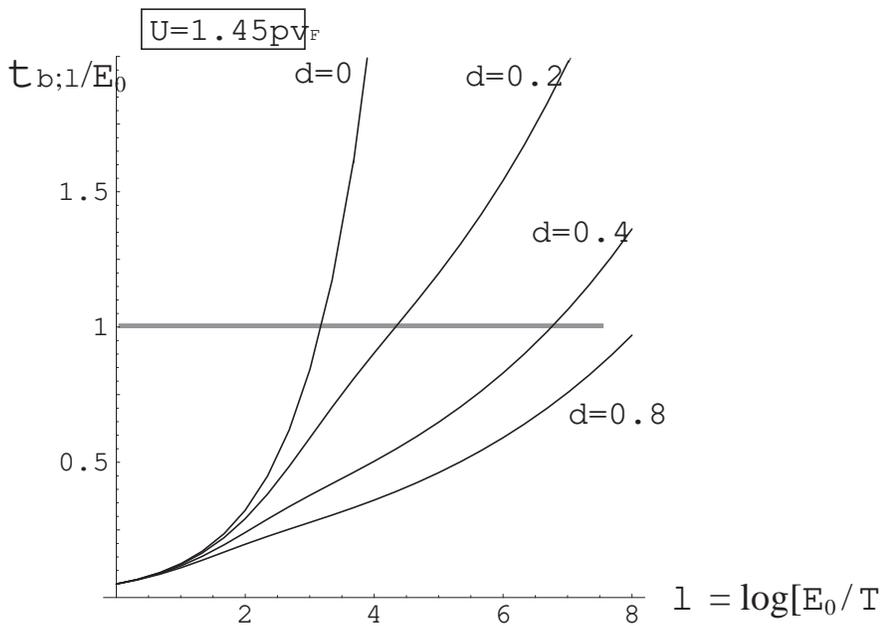
\begin{figure}
\caption{ RG flows of $t_{b;l}$ for   dimerization ratios, $\delta=0,0.2,0.4,0.8$.
}
\end{figure}

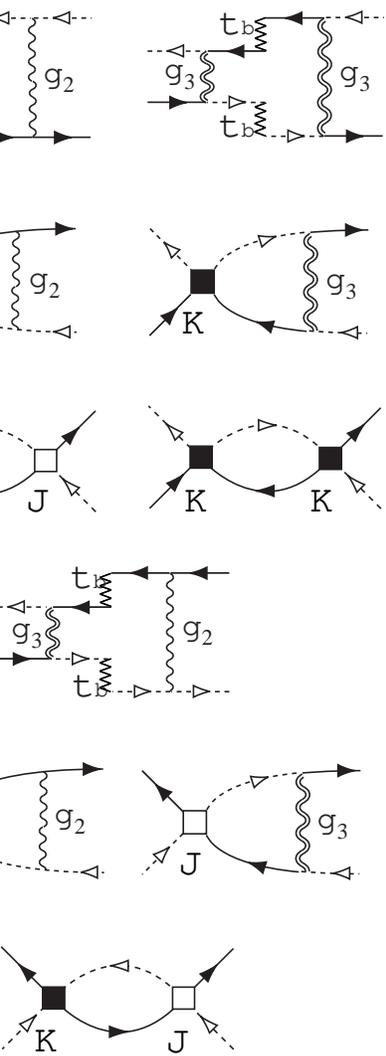
\begin{figure}
\caption{Renormalization of the ICEX processes in the AF channel for 
the (a) 
 normal and (b) umklapp processes.
}\end{figure}

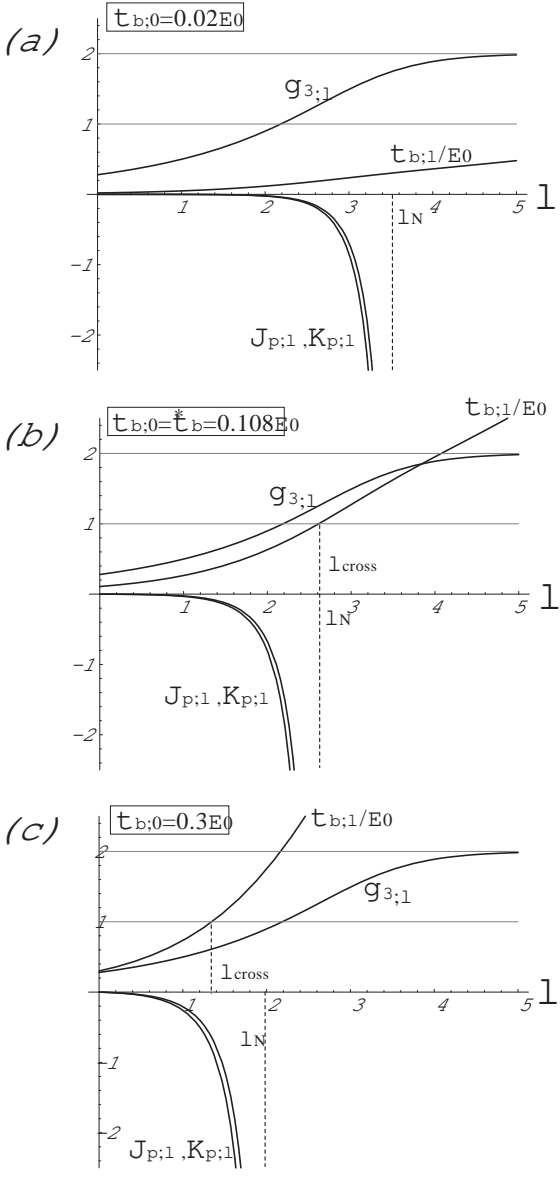
\begin{figure}
\caption{   RG flows of  $g_{3;l}$, $t_{b;l}/E_{0}$, 
$J_{q_b;l}$ and $K_{q_b;l}$ with $q_b=\pi$ for  $\delta=0.2$ and
$ t_{b0}/E_0=0.02, 0.108,   0.3$,
where  the vertical  lines show the locations of $l_{\rm cross}$ and $l_{N}$.
}
\end{figure}

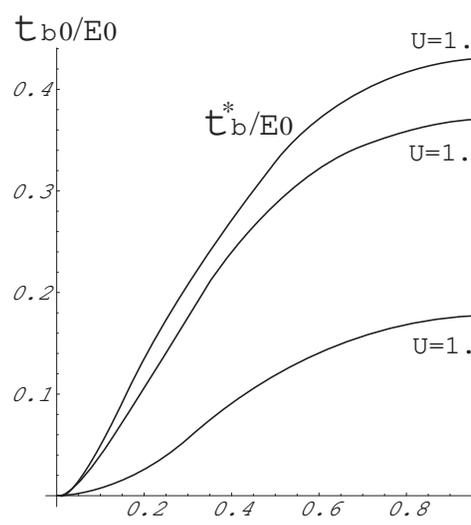
\begin{figure}
\caption{ Dependence of $t_{b}^{\ast}$  
  on the dimerization ratio $\delta$ for $U/\pi v_F=1.0,1.45,1.6$.
}
\end{figure}

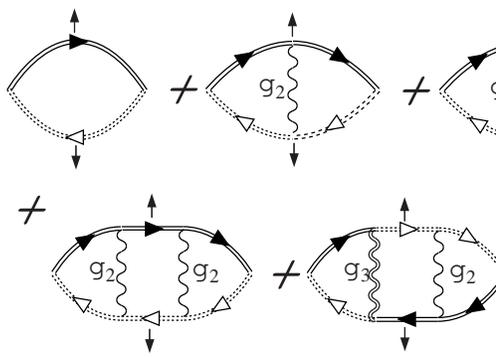
\begin{figure}
\caption{A series of diagrams representing 
the particle-hole fluctuations which contribute to the  transverse susceptibility within the RPA.
The double solid and broken lines represent
the two-dimensional one-particle propagators for the right- and left-moving sectors,
$
{\cal G}_{R}^{\rm 2D} 
$
and $
{\cal G}_{L}^{\rm 2D},
$
respectively.
The single and double wavy lines represent the intrachain $g$-ology interaction for the normal and umklapp processes,
respectively.
}\end{figure}

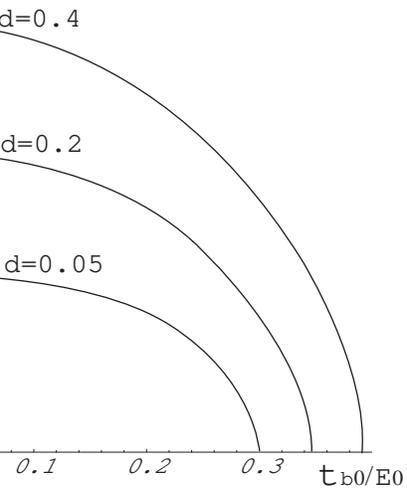
\begin{figure}
\caption{ Dependence of $T_{\rm N}^{\rm RPA}$ on $t_{b0}$ for $U=1.45\pi v_F$
and $\delta=0.05,0.2,0.4$.
}\end{figure}

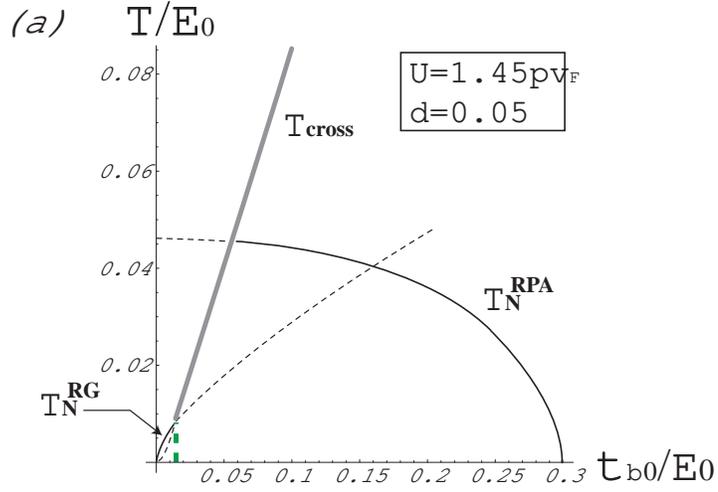
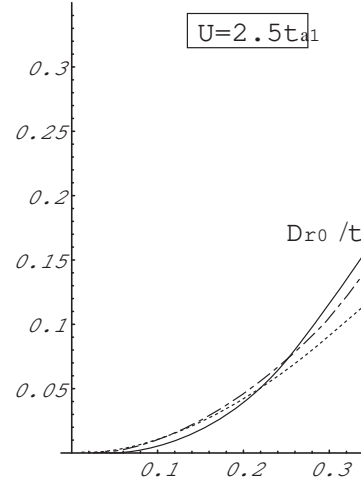
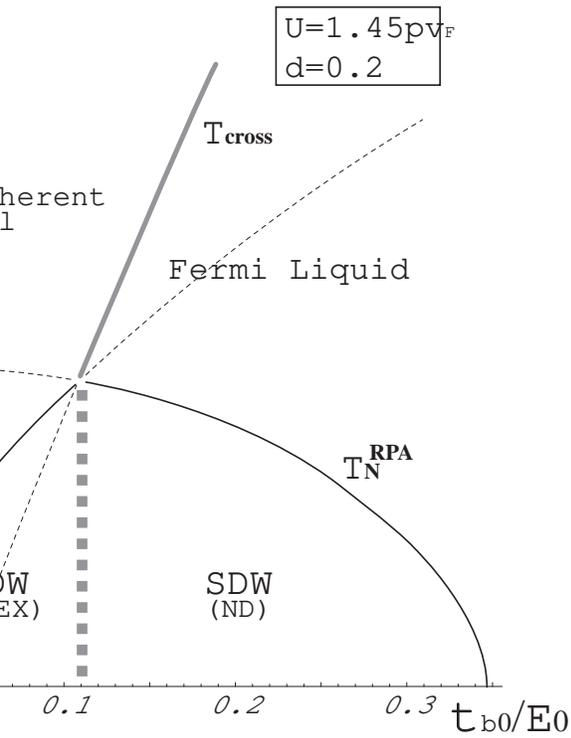
\begin{figure}
\caption{SDW phase diagram   for    $U=1.45\pi v_F$ and $\delta=0.2$.
{\bf SDW (ICEX)} stands for the spin density wave phase driven by
the interchain  exchange process in  the AF channel.
{\bf SDW (ND)} stands for the spin density wave phase driven by the Fermi surface nesting.
}
\end{figure}

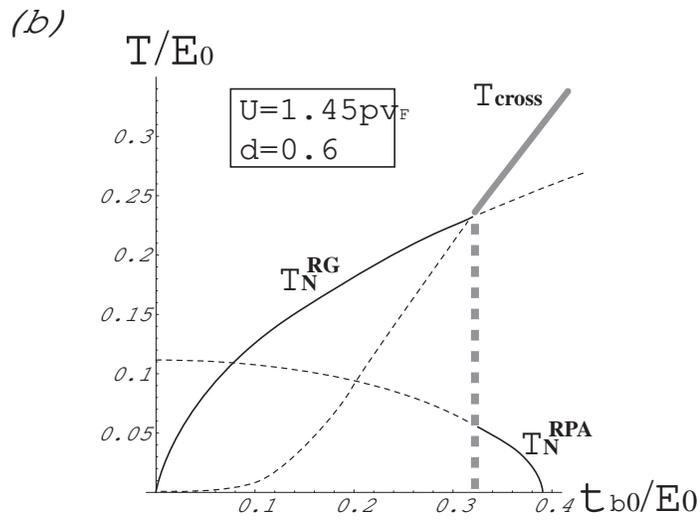
\begin{figure}
\caption{ SDW phase diagrams for      dimerization ratios 
(a) smaller ($\delta=0.05$)  or (b) larger ($\delta=0.6$) than that   of Fig.~11.
}
\end{figure}

\begin{figure}
\caption{ 
$\delta$-dependence of $\Delta_{\rho 0}/t_{a1}$, $t_{b}^\ast/t_{a1}$ and $T_{\rm um}/t_{a1}$  
in the case of $W=8E_0$ and $U=2.5t_{a1}$. 
}
\end{figure}

\begin{thebibliography}{999}
\bibitem{comment0}  Recall that the  1D non interacting retarded one-particle propagator  is given by ${\cal G}_R(x,t=0)\propto { 1 \over \xi  \sinh [x/\xi ] },$ which behaves as  $x^{-1}$ for $x\ll \xi$, while  $e^{-x/\xi}$ for $x\gg \xi$.
\end{thebibliography}
\end{document}